\documentclass[twocolumn,floatfix,prl,aps,showpacs]{revtex4-1}
\usepackage{graphicx,amsmath,amssymb}

\begin{document}

\title{Landau level mixing and the emergence of Pfaffian excitations for the 5/2 fractional quantum Hall effect}
\author{
   Arkadiusz W\'ojs$^{1,2}$, 
   Csaba T\H oke$^{3}$, and 
   Jainendra K.~Jain$^{4}$}
\affiliation{
   $^{1}$TCM Group, Cavendish Laboratory, 
   University of Cambridge,
   Cambridge CB3 0HE, United Kingdom}
\affiliation{
   $^{2}$Institute of Physics, 
   Wroclaw University of Technology,
   50-370 Wroclaw, Poland}
\affiliation{
   $^3$Institute of Physics, 
   University of P\'ecs, 
   7624 P\'ecs, Hungary}
\affiliation{
   $^{4}$Department of Physics, 
   104 Davey Lab, 
   Pennsylvania State University, 
   University Park PA, 16802}
\date{\today}

\begin{abstract}
We report on exact diagonalization studies for fully spin polarized 5/2 fractional quantum Hall effect, incorporating Landau level mixing through the Bishara-Nayak effective interaction. We find that there is an experimentally accessible region in the phase diagram where the Pfaffian model accurately describes not only the ground state but also the neutral and charged excitations. These results are consistent with the observed persistence of the 5/2 Hall effect down to very low magnetic fields; they are also relevant to the experimental attempts to detect nonabelian braid statistics. 
\end{abstract}


\maketitle

The proposed route to nonabelian braid statistics in the 5/2 fractional quantum Hall effect (FQHE) \cite{Moore91,Greiter91,Read00}, discovered more than two decades ago \cite{Willett87}, proceeds through a sequence of remarkable emergences. To minimize the bulk of the repulsive interaction, electrons in the second Landau level (LL) dress themselves with two vortices to transform into composite fermions \cite{Jain89}; composite fermions (CFs) experience a vanishing magnetic field and form a Fermi sea \cite{Halperin93}; the CF Fermi sea, however, is unstable due to a weak residual attractive interaction between composite fermions, which causes an equal spin $p_x\pm ip_y$ pairing, thereby opening a gap and producing an FQHE state \cite{Moore91,Greiter91,Scarola00}; the Abrikosov vortices of the paired CF state support zero mode solutions, namely majorana CFs, which are symmetric combinations of the CF creation and annihilation operators \cite{Read00}; these obey nonabelian braid statistics \cite{Read00} and are potential candidates for fault tolerant topological quantum computation \cite{DasSarma05}.

While the decisive verification of these ideas will eventually come from laboratory experiments, we have come to expect any successful theoretical postulate in the field of FQHE to pass the test against ``computer experiments,'' that is, exact solutions of the full many body problem for finite systems typically containing up to 16-18 particles. A concrete realization of the above physics is through the Pfaffian (Pf) model of Moore and Read \cite{Moore91}, which has been subjected to such tests. The Pf wave function for the ground state has a moderate overlap with the exact Coulomb ground state, which can be improved either by artificially strengthening the short range part of the interaction \cite{Morf98,Rezayi00,Moller08} or by considering the effect of finite thickness \cite{Peterson08}. The sensitivity to such slight modifications in the interaction indicates that the physical 5/2 FQHE state lies close to an instability. An adiabatic connection has been shown in finite system studies between the Coulomb and the Pf {\em ground} states \cite{Moller08,Storni10,Wojs09}. The situation is less clear for excitations, however. A test of the Pf quasiholes has not found them to be satisfactory approximations of the actual quasiholes of the unperturbed Coulomb interaction \cite{Toke07}, and an adiabatic connection between the Pf and the Coulomb quasiparticles and quasiholes has not yet been demonstrated. Given that the nonabelian braid properties of the quasiparticles and quasiholes are of primary interest, it would appear important to ascertain the region of validity of the Pf model for the excitations as well. Our results below provide strong evidence that sufficient amount of LL mixing produces not only a ground state very close to the Pf state, but also neutral and charged excitations that are well consistent with the Pf model. 

We employ the standard spherical geometry in our calculations, wherein $N$ electrons on the surface of the sphere are exposed to a magnetic flux of $2Q hc/e$, where $2Q$ is an integer. We consider electrons in the second Landau level coupled by the Bishara-Nayak (BN) effective interaction \cite{Bishara09} 
\begin{equation}\label{BN}
H_{\rm BN}=V_{\rm Coulomb}+\sum_m \delta {\cal V}_m+\sum_{m}{\cal W}_{m},
\end{equation}
where $V_{\rm Coulomb}$ is the second LL Coulomb interaction and the last two terms effectively account for perturbative corrections to the interaction due to LL mixing. They are defined as $\delta{\cal V}_m=\delta V_m\sum_{i<j}P^{(2)}_{ij}(2Q-m)$ and ${\cal W}_m=W_m\sum_{i<j<k}P^{(3)}_{ijk}(3Q-m)$, where $P^{(2)}_{ij}(L)$ and $P^{(3)}_{ijk}(L)$ project the state of the two and three particles into the subspace of total orbital angular momentum $L$, $\delta V_m$ is the change in the energy of a pair of electrons in relative angular momentum states $m$ ($m=1, 3, 5,\cdots$ for fully spin polarized electrons), and $W_{m}$ is the energy of a collection of three electrons with relative angular momenta $m=3, 5, 6, 7, 8\cdots$. The relevant expansion parameter in the perturbative calculation of Bishara and Nayak is the ratio of Coulomb energy scale to the cyclotron energy, $\kappa=(e^2/\epsilon \lambda)/(\hbar \omega_c)$, where $\lambda=\sqrt{\hbar c/eB}$ is the magnetic length and $\omega_c=eB/m_bc$ is the cyclotron frequency, $m_b$ being the band mass of the electron. For parameters appropriate to GaAs we have $\kappa\approx 2.5/\sqrt{B[T]}$ ($B[T]$ is magnetic field quoted in Tesla), which changes from $\kappa=0.8$ at $B=9$T to $\kappa=1.8$ at $B=2$T \cite{commentBN}. The values of $V_m$ and $W_m$ are given in Ref.~\onlinecite{Bishara09}. 

\begin{figure}[t]
\begin{center}
\includegraphics[width=\columnwidth]{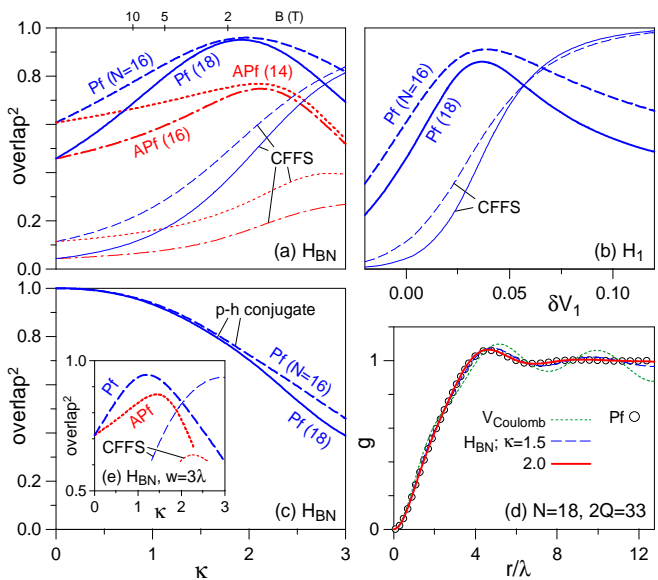}
\end{center}
\caption{\label{PfAPf} (Color online) Panel (a) shows the squared overlaps of the exact ground state of $H_{\rm BN}$ with both Pf and APf as a function of $\kappa$ for the indicated values of $N$. The top axis shows the corresponding magnetic field for GaAs. The squared overlaps of these states are also shown with the corresponding $L=0$ states of noninteracting composite fermions (lines with same color and style but thinner), labeled CFFS (CF Fermi sea \cite{commentCF}). Panel (b) shows the squared overlaps of the exact ground state of $H_1$ with both Pf and CFFS. Panel (c) displays the squared overlap between the ground state of $H_{\rm BN}$ at $2Q=2N-3$ with the hole conjugate of its ground state at $2Q=2N+1$; the deviation from unity is a measure of the p-h symmetry breaking. The pair correlation function of the $H_{\rm BN}$ ground state is shown in panel (d) for several values of $\kappa$ (with $\kappa=0$ giving the Coulomb result) along with that of the Pfaffian wave function. Panel (e) depicts comparisons for Pf and APf ($N=16$ and 14) with the $H_{\rm BN}$ ground state for a quantum well of thickness $w=3\lambda$ (three magnetic lengths).
} 
\end{figure}

The solution of the short range three body interaction $H_{\rm Pf}=W_3$ {\em defines} the Pf model in what follows. The single zero energy eigenstate of $H_{\rm Pf}$ at $2Q=2N-3$ is identified as the Pf ground state, the zero energy solutions at $2Q>2N-3$ as Pf quasihole (QH) states, and the low energy solutions at $2Q<2N-3$ as Pf quasiparticle (QP) states. As another reference we also consider the two body interaction $H_1=V_{\rm Coulomb}+\delta V_1$, wherein we arbitrarily modify the two body pseudopotential $V_1$.

A fundamental aspect of the Pf state is its lack of particle-hole (p-h) symmetry: the Pf is topologically distinct from its hole partner, called the antiPfaffian (APf) \cite{Levin07}. In the absence of LL mixing, the Pf and APf states are degenerate in the thermodynamic limit, and LL mixing, which enters through a three body interaction, will split the degeneracy to favor one over the other \cite{Bishara09,Peterson08b,Wang09}. (The two have different triplet amplitudes, which guarantees an extensive energy difference in the thermodynamic limit.) In a recent work Rezayi and Simon \cite{Rezayi10} have argued that the APf has lower energy. Recent tunneling experiments \cite{Radu08} are closer to the expectation from the APf state, although not conclusive. 

We first determine which state is favored by the BN interaction, which is different from the one used in Ref.~\cite{Rezayi10}. Fig.~\ref{PfAPf}(a) demonstrates that increasing $\kappa$, which breaks p-h symmetry, produces a ground state that is very well approximated by the Pf wave function; the squared overlaps between the exact $H_{\rm BN}$ ground state and the Pf increase from 0.4-0.6 at $\kappa=0$ to above 0.95 at $\kappa=2$ for systems with up to 18 particles. Furthermore, the Pf is a significantly better approximation than the APf, indicating that $H_{\rm BN}$ selects the Pf. An extrapolation of the $H_{\rm BN}$ energies of the Pf and the APf states (not shown) is also consistent with that conclusion. Our calculations thus indicate that the BN interaction selects the Pf phase. Further work will be required to ascertain which model is most reliable for LL mixing and which state is stabilized under realistic conditions, but we will focus on the Pf phase in what follows. We also note that another earlier work in this direction \cite{Wojs06} used a different approach to account for LL mixing, by enlarging the Hilbert space to allow a single excited electron; it considered only the APf, and was restricted, because of the significantly enlarged Hilbert space, to very small systems (8 particles). 

Fig.~\ref{PfAPf}(b) reminds that increasing the pseudopotential $V_1$ also produces high overlaps with the Pf wave function (also followed by a transition into the CF Fermi sea, consistent with the interpretation of the Pf as a weakly paired state of composite fermions) \cite{Rezayi00,Moller08}. The increase in $V_1$ is not microscopically motivated, however, and also does not lift the degeneracy between the Pf and the APf.

\begin{figure}[t]
\begin{center}
\includegraphics[width=\columnwidth]{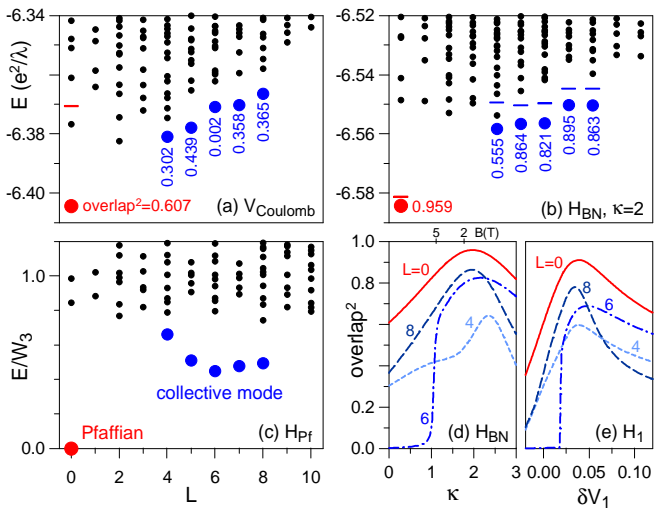}
\end{center}
\caption{\label{GS} (Color online) Spectra of 16 particles at flux $2Q=2N-3=29$ for (a) $V_{\rm Coulomb}$, (b) $H_{\rm BN}$, and (c) $H_{\rm Pf}$ (black as well as colored dots). The energy expectation values of the Pf wave functions are also shown for $V_{\rm Coulomb}$ and $H_{\rm BN}$ in panels (a) and (b) (dashes) except when they are so high that they fall outside the frame. The energies are given in units of $e^2/\lambda$ ($\lambda$ is the magnetic length) in panels (a) and (b) and in units of $W_3$ in (c). The numbers near the ground state (red) and the low energy neutral excitations (blue) indicate the squared overlaps with the corresponding Pf eigenstates. Panels (d) and (e) show the evolution of the overlaps of the exact states of $H_{\rm BN}$ and $H_1$ with the Pf states as a function of $\kappa$ and $\delta V_1$, respectively; results for only alternate $L$ are shown for clarity.
} 
\end{figure}

\begin{figure}[t]
\begin{center}
\includegraphics[width=\columnwidth]{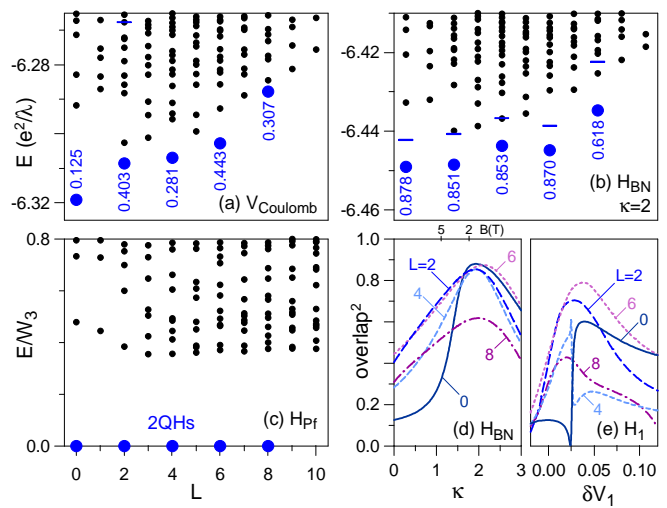}
\end{center}
\caption{\label{2QH} (Color online) Same as in Fig.~\ref{GS} but for two Pf quasiholes at $2Q=2N-2=30$. 
} 
\end{figure}

\begin{figure}[t]
\begin{center}
\includegraphics[width=\columnwidth]{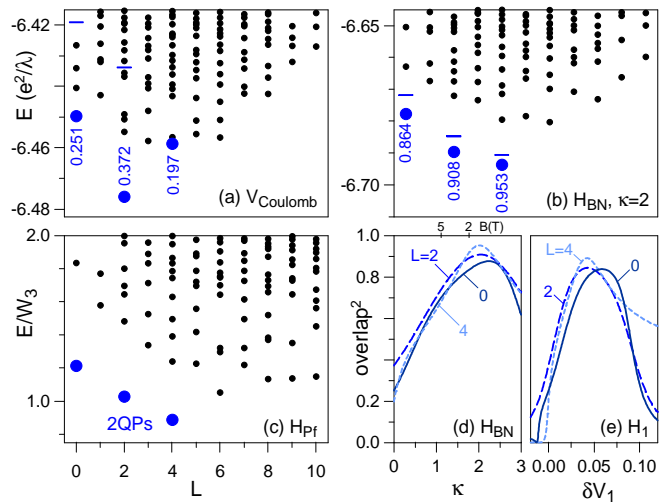}
\end{center}
\caption{\label{2QP} (Color online) Same as in Fig.~\ref{GS} but for two Pf quasiparticles at $2Q=2N-4=28$.
} 
\end{figure}

The next three figures compare the neutral and charged excitations of $H_{\rm BN}$ with the corresponding Pf excitations. Fig.~\ref{GS} shows the spectra at $2Q=2N-3$ for the Coulomb, Pf, and BN interactions (the last for $\kappa=2$), as well as the overlaps as a function of $\kappa$ and $\delta V_1$. Figs.~\ref{2QH} and \ref{2QP} show similar comparisons for two quasiholes at $2Q=2N-2$ and two quasiparticles at $2Q=2N-4$. The Pf model $H_{\rm Pf}$ produces a zero energy ground state in Fig.~\ref{GS}(c), as well as a well defined neutral exciton branch, which presumably represents a QP-QH excitation; it also produces a zero energy Pf QH band in Fig.~\ref{2QH}(c), and a low energy band in Fig.~\ref{2QP}(c) that we identify as the Pf QP states. A comparison with the exact eigenstates of the BN interaction brings out the following. First of all, the neutral and the charged excitations of $V_{\rm Coulomb}$ (panel a) are not well described by the Pf model, as indicated by the overlap values as well as the Coulomb energies of the Pf eigenstates which often lie outside the frame. With LL mixing, however, the agreement improves rapidly (panels b and d). The overlaps become much larger, and the expectation values of $H_{\rm BN}$ with respect to the Pf eigenstates of $H_{\rm Pf}$ (shown as dashes) are in qualitative and semi-quantitative agreement with the actual band. The somewhat worse agreement at the smallest $L$ in the exciton branch in Fig.~\ref{GS} and at the largest $L$ in Fig.~\ref{2QH} suggests that the Pf model is less accurate for short distance physics, because the Pf QH and the Pf QP in the exciton branch are at their closest separation at the smallest angular momentum ($L=4$) and the two QHs are nearest at the largest $L$ in Fig.~\ref{2QH}. These comparisons demonstrate that, with LL mixing, the Pf physics emerges for quasiparticles and quasiholes. (Results are shown here only for the largest systems that we have studied, but those from smaller systems are also fully consistent.)  

\begin{figure}[t]
\begin{center}
\includegraphics[width=\columnwidth]{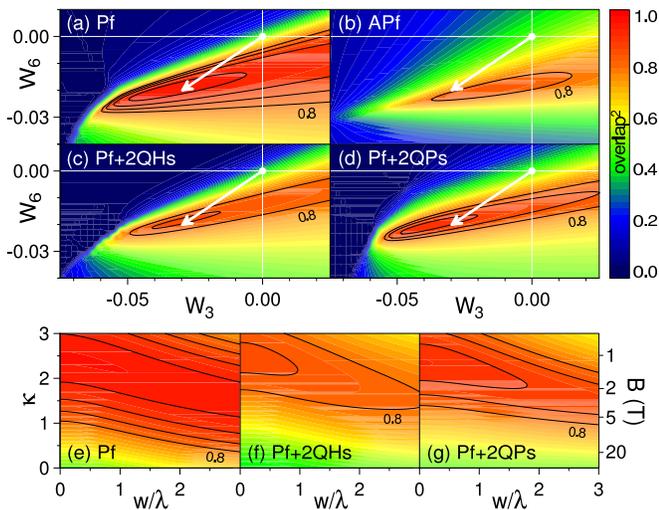}
\end{center}
\caption{\label{fig5} (Color online) Squared overlaps of the Pf ground and QP/QH states with the exact eigenstates as a function of interaction parameters. Panels (a) and (e) are for the Pf ground state; (c) and (f) for Pf+2QHs; (d) and (g) for Pf+2QPs. All of these are for $N=14$ particles, with $L=0$ in (a) and (e) and $L=1$ in (c), (d), (f) and (g) which corresponds to maximally separated QPs or QHs. The comparison with the APf is shown in panel (b) for 12 particles. In (a)-(d), the results are given as a function of the 3-body pseudopotentials $W_3$ and $W_6$, with $W_5=0.367\,W_3$; the white arrows trace the BN interaction, with the two ends representing $\kappa=0$ and $2$. Panels (e)-(g) show the results as a function of $\kappa$ and the quantum well width $w$, quoted in units of the magnetic length $\lambda$. The contours mark squared overlaps of 0.8, 0.85, 0.9 and 0.95. 
} 
\end{figure}

To investigate the robustness of this physics to the precise form of the interaction, we have studied the effect of a variation of the pseudopotentials. For ease of presentation, we set $\delta {\cal V}_m=0$ and retain only the dominant three body pseudopotentials of the BN model. In Fig.~\ref{fig5} we vary $W_3$ and $W_6$, and take $W_5=0.367\,W_3$ (as in the BN interaction). We can conclude that there is a range of parameters where the Pf physics is valid. The slightly worse agreement for Pf QHs in this figure as well as in Fig.~\ref{2QH} can be attributed to the closer proximity of this system to the APf, leading to a stronger interference with the APf physics. This, however, is a finite size effect and should not be relevant in the thermodynamic limit where either the Pf or the APf ground state would be chosen.

The importance of finite quantum well width $w$ has been stressed previously \cite{Peterson08}. To assess the role of LL mixing for a system with finite thickness, we have studied an approximate model in which we modify $V_{\rm Coulomb}$ (assuming infinite square quantum well confinement) and rescale $\delta{\cal V}$ and $\delta{\cal W}$ by the ratio $V_{\rm Coulomb}(m=1,w)/V_{\rm Coulomb}(m=1,w=0)$ (which is 0.91 for $w=3 \lambda$). The optimal values of $\kappa$ shift downward with increasing thickness, as seen in Figs.~\ref{PfAPf}(e) and \ref{fig5}(e)-(g). We note that, in the absence of LL mixing, the Pf QH and Pf QP states do not improve as rapidly with increasing thickness as does the Pf ground state. 

The assumption of full spin polarization is not always justified. The 5/2 FQHE has been seen at a moderately high field of $B=10$~T \cite{Zhang10}, where the state is expected to be fully spin polarized, but also at very low fields \cite{Dean08}. There is numerical evidence that the 5/2 state remains fully polarized even at low fields \cite{Morf98,Rezayi10,Park98}. Inelastic light scattering experiments suggest a lack of full spin polarization at somewhat elevated temperatures \cite{Pinczuk}, but do not directly probe the 5/2 FQHE state, and can be explained \cite{Wojs10} in terms of a polarized ground state with disorder aided depolarization due to $e/2$-charged skyrmions. We have not considered the effect of disorder \cite{Morf03}.

LL mixing typically results in a reduction of the gap. However, due to the strengthening of the Pf physics, the 5/2 gap in our calculations slightly increases until $\kappa\approx 1.5$ before it begins to decrease. This is nicely consistent with the observation of a robust 5/2 FQHE at very low magnetic fields ($\sim2.5$~T) \cite{Dean08}. Fig.~\ref{fig5}(f)-(g) suggests the possibility that LL mixing might actually be essential for establishing the Pf physics for the quasiparticles and quasiholes, but further theoretical and experimental work will be needed to settle this issue. 

In summary, our principal result is that there exists a realistic Hamiltonian, including LL mixing, for which the Pf physics is demonstrably established not only for the ground state but for the excitations as well. 

We thank N. R. Cooper, S. Das Sarma, F. D. M. Haldane, T. Jolicoeur, G. M\"oller, E. H. Rezayi, and S. H. Simon for useful comments, and acknowledge support by the Marie Curie grant PIEF-GA-2008-221701 (A.W.) and Science, Please! Innovative Research Teams, SROP-4.2.2/08/1/2008-0011 (C.T.).


\begin{thebibliography}{99}
\bibitem{commentBN} The Bishara-Nayak interaction $H_{\rm BN}$ was derived perturbatively, and is therefore strictly valid only for small $\kappa$; however we will assume it here for large $\kappa$ as well.
\bibitem{commentCF} We define here the Coulomb eigenstates in the lowest LL as the weakly interacting CF states; the two are known to be almost identical.
\end{thebibliography}
\end{document}